\documentstyle[aps,preprint,epsf]{revtex}

\def\I{{\cal I}}
\def\M{{\cal M}}
\def\S{{\cal S}}
\def\d{{\rm d}}
\def\to{\rightarrow}
\def\as{\alpha_s}
\def\dc{d_{\rm cut}}
\def\yc{y_{\rm cut}}
\def\MeV{{\rm MeV}}
\def\GeV{{\rm GeV}}
\newcommand{\la}{\langle}
\newcommand{\ra}{\rangle}
\newcommand{\Nf}{N_f}
\def\NLO{next-to-leading order }
\def\del#1{\lower.25em\hbox{\LARGE $\times$}\kern -1em #1 }

\newcommand{\beq}{\begin{equation}}
\newcommand{\eeq}{\end{equation}}
\newcommand{\beqn}{\begin{eqnarray}}
\newcommand{\eeqn}{\end{eqnarray}}

\def\nn{\nonumber}
\def\ms{{\overline{{\rm MS}}}}

\begin{document}

\begin{titlepage}
\vspace*{-2cm}
\begin{flushright}
KLTE-DTP/96-2\\
April, 1996 \\
\end{flushright}
\vskip .5in
\begin{center}
{\large\bf
Three-jet cross section in hadron collisions at next-to-leading order:\\
pure gluon processes}\\
\vskip 0.5cm 
Zolt\'an Tr\'ocs\'anyi\footnote{Zolt\'an Magyary fellow} \\
\vskip 0.2cm
\it Department of Theoretical Physics, KLTE, Debrecen, Hungary \\
\end{center}

\begin{abstract}
\noindent 
The \NLO three-jet cross section in hadron collisions is calculated in
the simplified case when the matrix elements of all QCD subprocesses are
approximated by the pure gluon matrix element. The
longitudinally-invariant $k_\perp$ jet-clustering algorithm is used. The
important property of reduced renormalization and factorization scale
dependence of the \NLO physical cross section as compared to the Born
cross section is demonstrated.
\bigskip

\end{abstract}
\end{titlepage}
\setcounter{footnote}{0}

The structure of hadronic final states in high energy collisions may be
described in terms of jet characteristics. Nowadays, jet cross section
data are used both for precise quantitative tests of QCD --- such as
measurement of the strong coupling $\as$ and QCD scale $\Lambda_\ms$ ---
and looking for signs of new physics beyond the standard model. 
Both of these objectives have been studied extensively in $e^+e^-$
annihilation \cite{WebGlas} that is characterized by a fixed high energy
scale, namely the machine energy. One would also like to analyze the
data of scattering processes at the highest available energy scale,
where the signs of new physics is expected to be the most profound. The
highest scale is in general found in hadron collisions. Unfortunately,
there are important ambiguities which limit our ability to perform high
precision quantitative studies with the jet cross sections observed in
hadron collisions. In experiments ambiguities arise from the question of
how to define a jet and from the systematic uncertainties of jet energy
measurements. On the theoretical side, apart from the ubiquitous and ever
decreasing uncertainty in the parton density functions \cite{parton},
there still remains uncertainty in the choice of renormalization and
factorization scales, the magnitude of the higher order corrections and
the question of how to match the theoretical and experimental jet
definitions. The theoretical ambiguity coming from these points can be
decreased if the \NLO corrections are calculated.

The most easily calculated \NLO corrections to cross sections in hadron
collisions are those to the inclusive
one-jet and two-jet cross sections which have been available
for some time \cite{EKSinclg,onetwojet}. These results have already
proven to be extremely important: the significant enhancement in the
experimental one-jet cross section over the result of the theory in the
range $p_T^{\rm jet}>200\,\GeV$ may be interpreted as signal of physics
beyond the Standard Model \cite{CDFqstruct}.

In order to be able to calculate \NLO corrections for more complex final
states than the ones mentioned above --- such as the three-jet cross
section in hadron collisions ---, two obstacles had to be overcome.
Firstly, there was the issue of loop matrix elements which only
recently have become available for all five-parton processes
\cite{BDK5g,KST4q1g,BDK2q3g} necessary for a three-jet analysis.
Secondly, the algorithm for the cancelation of infrared divergences
applied in ref.\ \cite{onetwojet} were not directly applicable in
processes with more complex final state kinematics. Recently, a number
of general schemes have been proposed for achieving the cancelation of
final state infrared singularities and mass
factorization both in the framework of ``phase space slicing'' method
\cite{GGKpp} and ``subtraction'' method \cite{FKSjets,CSdipol,NTjets}.

In this letter we present a brief summary of an analysis of three-jet
cross sections using the algorithm of ref.\ \cite{NTjets}, but in the
simplified  case when all squared matrix elements are approximated with
that of the pure gluon subprocess. Thus the results shown are intended
only for demonstrating the applicability of the subtraction scheme of
ref.\ \cite{NTjets} in the case of hadron collisions rather than a
serious theoretical description of the data. We anticipate however,
similar conclusions as those drawn here will hold once the complete
analysis with quarks included is finished.

According to the factorization theorem, the \NLO infrared safe physical
cross section at order $\as^{(N+1)}$ is a sum of two
integrals,
\beq
\label{physical}
\sigma = \I[2\to N] + \I[2\to N+1],
\eeq
where in the case of the pure gluon approximation for the squared matrix
element, these integrals have the form
\beqn
\label{I2ton}
&&\I[2\to n] =
\int \d x_A f_{\rm eff} (g,x_A)\int \d x_B f_{\rm eff} (g,x_B)
\\ \nn &&\qquad\qquad\times  
\frac{1}{2 x_A x_B s}\frac{1}{n!} \int \d \Gamma^{(n)}(p_1^\mu,\ldots,p_n^\mu)
\la |\M (g+g\to n g)|^2 \ra \S_n(p_1^\mu,\ldots,p_n^\mu).
\eeqn
In this equation $\d \Gamma^{(n)}$ is the usual
$n$-particle phase space measure. There are two possibilities for the
choice of the effective gluon densities. One can either imagine
colliding glueballs \cite{EKSinclg}, use a reference gluon density
at a fixed $\mu_0$ scale and evolve it to other scales with $\Nf=0$,
or alternatively, one can use the effective gluon density \cite{esfag},
\beq
f_{\rm eff}(g,x) = f(g,x)+\frac{4}{9}\sum_q[f(q,x)+f(\bar{q},x)].
\eeq
Our choice will be the latter one. There is still the question of what
sort of strong coupling $\as(\mu)$ one should use in the pure
gluon $\la |\M (g+g\to n g)|^2 \ra$ squared matrix elements. We 
use the two-loop formula for the strong coupling with $\Nf=0$ and
the QCD scale parameter in the modified minimal subtraction scheme,
$\Lambda_\ms$ chosen to be $1600\,\MeV$ so as $\as(50\,\GeV)\simeq 0.13$
is consistent with the value of $\as(50\,\GeV)$ observed in the real
world with quarks. In this way we ensure that the relative size of the
\NLO correction is similar to that in the full QCD case. Note however,
when we compare the order $\as^{(N+1)}$ cross section to the results of
the Born-level calculation, we compute the Born cross section using the
one-loop formula for $\as$ with the $\Lambda_\ms=1100\,\MeV$, which makes
$\as(50\,\GeV)$ about 15\,\% and the Born-level three-jet cross section
about 50\,\% bigger.

The function $S_n$ in eq.\ (\ref{I2ton}) represents the physical quantity
to be calculated.  Among the numerous infrared safe physical quantities
one can calculate with the technique presented ref.\ \cite{NTjets}, an
explicit example, that we use in the present analysis, is the \NLO
three-jet cross section in hadron collisions defined using the
longitudinally-invariant $k_\perp$ jet-clustering algorithm \cite{Catanietal}.
For hadron collisions the jet-clustering algorithm is a two-stage process,
each characterized by a scale. The first step is the pre-clustering of hadrons
into hard final state jets and beam jets. One sets the hardness scale of
the jets to $\dc$. Then for every final state hadron $h_k$ and for every 
pair $h_k$, $h_l$ one computes the corresponding value of the resolution
variables $d_{kB}$ and $d_{kl}$. There are several possibilities for the
definition of the resolution variables. For instance, we may choose
\beq
d^2_{kB} = p_k^2,\quad d^2_{kl} = \min(p_k^2, p_l^2) R^2_{kl},
\eeq
where $R_{kl}$ is the distance in $(y,\phi)$-space,
\beq
R_{kl} = \sqrt{(\eta_k-\eta_l)^2 + (\phi_k-\phi_l)^2}
\eeq
and $(p_k,\theta_k,\phi_k)$ are the cylindrical coordinates of the 
three-momentum of hadron $h_k$ with respect to the beam axis, with 
$\eta_k=-\ln \tan(\theta_k/2)$ being the corresponding pseudorapidity.
(For other possibilities see ref.\ \cite{Catanietal}). Having calculated
the resolution variables, one considers the smallest value among
$\{d_{kB},d_{kl}\}$. If $d_{ij}$ is the smallest value and $d_{ij}<\dc$,
then $h_i$ and $h_j$ are combined into a single cluster with momentum
$p_{(ij)}^\mu$ according to a recombination prescription and the
algorithm starts again with hadrons $h_i$ and $h_j$ deleted from the
final state and the `pseudoparticle' of momentum $p_{(ij)}^\mu$ added to
the final state. If $d_{iB}$ is the smallest value and $d_{iB}<\dc$,
then hadron $h_i$ is deleted from the final state and is included in the
beam jets and the algorithm starts again. The algorithm stops if the
smallest value is larger than the hardness scale $\dc$. At the end of the
algorithm one has two beam jets and several hard final state jets.
The second step of the algorithm is the resolution of the event structure
into sub-jets. For this step one defines a resolution parameter $\yc$,
$\yc=Q_0^2/\dc^2\le 1$. Using this resolution parameter and the set of 
final state hadron momenta pre-clustered into the hard final state jets 
one performs a $k_\perp$ jet-clustering algorithm already familiar from
studies in $e^+e^-$ annihilation. For the sake of simplicity, here we
choose $\yc=1$. With this choice we focus our attention to hard final 
state jets only, and suppress the second step of the clustering
algorithm.

The principle of parton-hadron duality implies that we use the same
clustering algorithm at parton level as defined at hadron level. Thus
the measurement functions used for the \NLO perturbative calculation
of $N$ hard final state jet production are
\beq
\label{SN+1}
\S_{N+1}(p_1^\mu,\ldots,p_{N+1}^\mu)
=\Theta(d_{{\rm min}}^{(N+1)}>\dc)
+\Theta(d_{{\rm min}}^{(N+1)}<\dc)\Theta(d_{{\rm min}}^{(N+1\to N)}>\dc)
\eeq
and
\beq
\S_N(p_1^\mu,\ldots,p_N^\mu)
=\Theta(d_{{\rm min}}^{(N)}>\dc)
\eeq
where 
\beq
d_{{\rm min}}^{(n)}= \min(\{p_i^2,d_{ij}\}:\;i,j=1,\ldots,n,\;i\ne j),
\eeq
and $d_{{\rm min}}^{(N+1\to N)}$ is the minimal value of the resolution
variables after one clustering step.
In eq.\ (\ref{SN+1}) the first term represents the $N+1$-jet production,
while the second one represents the production of $N$ jets and either
a soft parton, or a hard parton collinear with the beam axis, thus
included in the beam jets or $N$-jet production such that all final 
state partons are hard, but two of them are collinear, thus combined 
into a single jet.
It is not difficult to check that these measurement functions are
infrared safe if any sensible recombination scheme \cite{Catanietal}
is applied. In our analysis, we use the $p_t$-weighted recombination.
In this scheme the transverse momentum, pseudorapidity and azimuth of
the pseudoparticle $(ij)$ are defined as
\beqn
p_{t(ij)}&=& p_{ti}+p_{tj}, \\
\eta_{t(ij)}&=& \frac{p_{ti}\eta_{ti}+p_{tj}\eta_{tj}}{p_{t(ij)}}, \\
\phi_{t(ij)}&=& \frac{p_{ti}\phi_{ti}+p_{tj}\phi_{tj}}{p_{t(ij)}}.
\eeqn
We remark that any other experimental cut, such as cut in the rapidity
window, or a $p_t$ trigger should also be included in the measurement
functions. In our analysis, we required that for the rapidity of jets
$|\eta|<3$ and the sum of the transverse momenta of the observed particles
$p_t^{\rm sum}>100\,\GeV$.

We now turn to the description of our results which were obtained at
$\sqrt{s}=1800\,\GeV$ machine energy and using the HMRS(B) \cite{hmrsb}
parton distributions. In Fig.~1 we plot the
total three-jet cross section both at Born level and at \NLO for a fixed
value of $\dc=70\,\GeV$ vs $\mu$ which is the common renormalization and
factorization scale. This plot demonstrates that over a wide range of
$\mu$ values the theoretical uncertainty coming from the scale dependence
is sizeably reduced in the \NLO result as compared to the result of a
Born calculation. In particular, one expects on general grounds that
$\mu$ should be chosen of the order of the hard scale $\dc$. If one
varies $\mu$ in the range of $\dc/2<\mu<2\dc$, then the change in the
\NLO cross section is less than 1/6 of the change in the Born-level
result. Thus the inclusion of the higher order correction decreases the
uncertainty in the theoretical prediction by a factor of larger than 6.
Similar conclusion can be drawn from plots at other $\dc$ values in
the range of $20\,\GeV<\dc<200\,\GeV$. This can also be seen from
Fig.~2, where the overall size of the three-jet cross section can be 
read off from a differential cross section $\dc^3\d\sigma/\d\dc$ plotted
vs $\dc$. The wide gray band shows the result of a Born level calculation
with $\mu$ varied between $\dc/2<\mu<2\dc$, while the narrower black band
inside is the \NLO result with the same scale variation.

In order to give the reader some feeling about the errors of the Monte
Carlo integrations, in Fig.~3 we plot the size of the Born-level cross
section and the higher order correction to it together with the statistical
errors of these results at $\mu=\dc$. The statistical error of the Born
result is below 1\,\%, while the statistical error of the full
\NLO cross section plotted in Fig.~2 is below 10\,\%.

In conclusion, we have calculated the three-jet cross section for the
longitudinally invariant $k_\perp$ jet-clustering algorithm in hadron
collisions for the simplified case when the matrix elements of all
subprocesses are approximated by those of the pure gluon subprocess.
The motivation for the particular choice of the jet definition is 
simply the pleasant property of the clustering algorithm that it uniquely
assinges all final state particles to a certain jet. Using this
definition one avoids the problem of jet separation in case of
overlapping jets that occurs when cone jet definition is used.
We used the subtraction method of ref.\ \cite{NTjets} for canceling the
infrared singularities.
This method has the important feature that the physical quantity to
be calculated is well separated from the theoretical problems of
cancelation of infrared singularities and can easily be changed in a
modular fashion in the Monte Carlo program. Thus the particular choice
for the jet definition is by no means essential. We have found that the
inclusion of the higher order correction improves dramatically our
theoretical description of the three-jet cross section: the ambiguity
coming from the arbitrary choice of the renormalization and factorization
scales is reduced by a factor of at least six.  Although the current
analysis is not complete in the sense that we have not used the full
QCD matrix elements, we anticipate similar conclusions once the effect
of quarks is included.

This research was supported in part by the EEC Programme "Human Capital
and Mobility", Network "Physics at High Energy Colliders", contract
PECO ERBCIPDCT 94 0613 as well as by the Foundation for Hungarian Higher 
education and Research, the Hungarian National Science Foundation grant
OTKA T-016613 and the Research Group in Physics of the Hungarian Academy
of Sciences, Debrecen.

\def\np#1#2#3  {Nucl.\ Phys.\ {\bf #1}, #2 (19#3)}
\def\pl#1#2#3  {Phys.\ Lett.\ {\bf #1}, #2 (19#3)}
\def\prep#1#2#3  {Phys.\ Rep.\ {\bf #1}, #2 (19#3)}
\def\prd#1#2#3 {Phys.\ Rev.\ D {\bf #1}, #2 (19#3)}
\def\prl#1#2#3 {Phys.\ Rev.\ Lett.\ {\bf #1}, #2 (19#3)}
\def\zpc#1#2#3  {Zeit.\ Phys.\ C {\bf #1}, #2 (19#3)}
\def\cmc#1#2#3  {Comp.\ Phys.\ Comm.\ {\bf #1}, #2 (19#3)}

%
%

\setcounter{figure}{0}
\begin{figure}
\epsfxsize=18cm \epsfbox{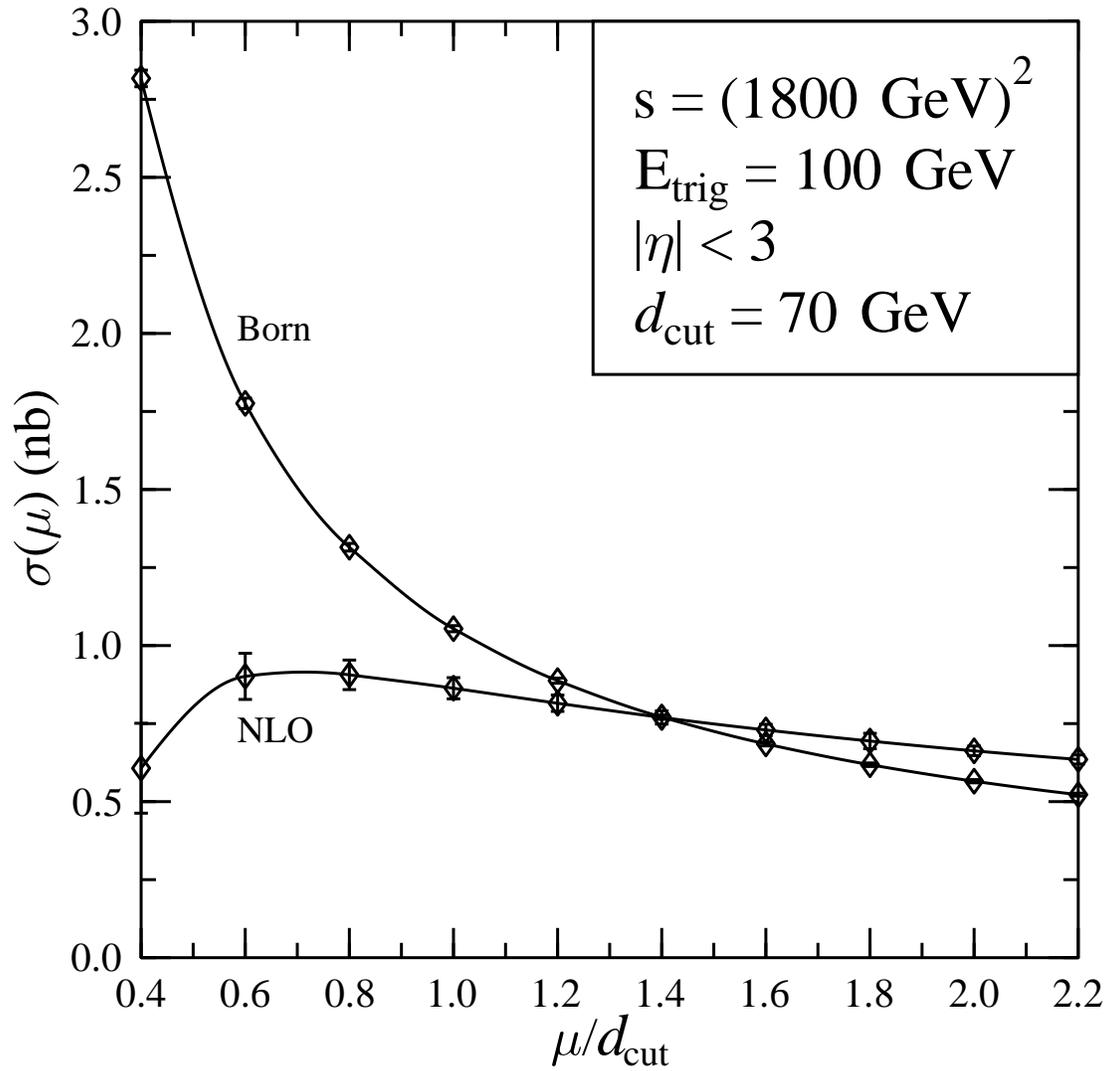}
\caption{Total three-jet cross section $\sigma(\dc=70\,{\rm GeV})$
vs $\mu$ at the Born and $\alpha_s^4$ level.}
\end{figure}

\begin{figure}
\epsfxsize=18cm \epsfbox{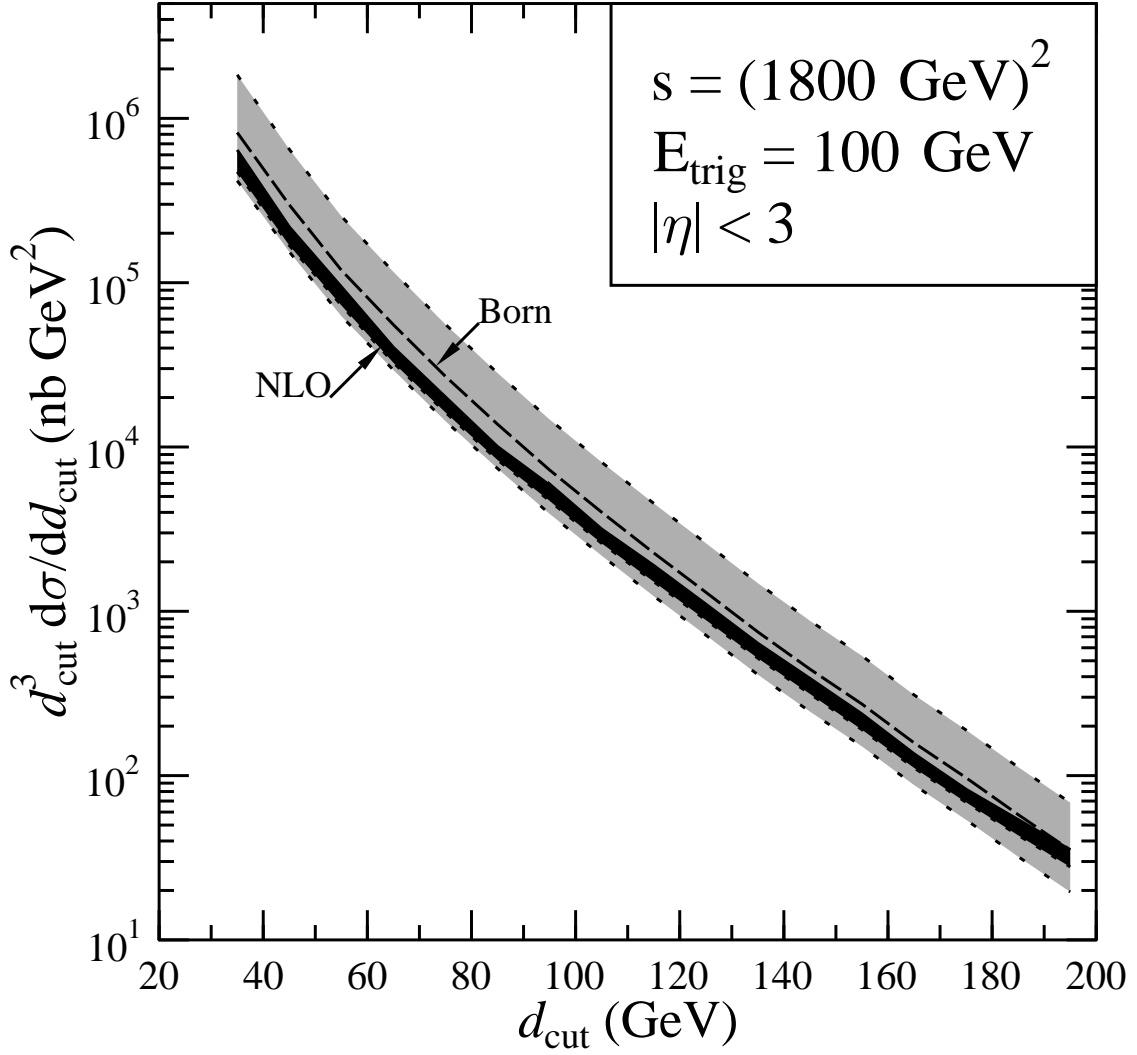}
\caption{Differential three-jet cross section $\dc^3\d\sigma/\d\dc$
vs $\dc$ for $0.5\dc<\mu<2\dc$ at Born level (gray band) and at
next-to-leading order (black band).}
\end{figure}

\begin{figure}
\epsfxsize=18cm \epsfbox{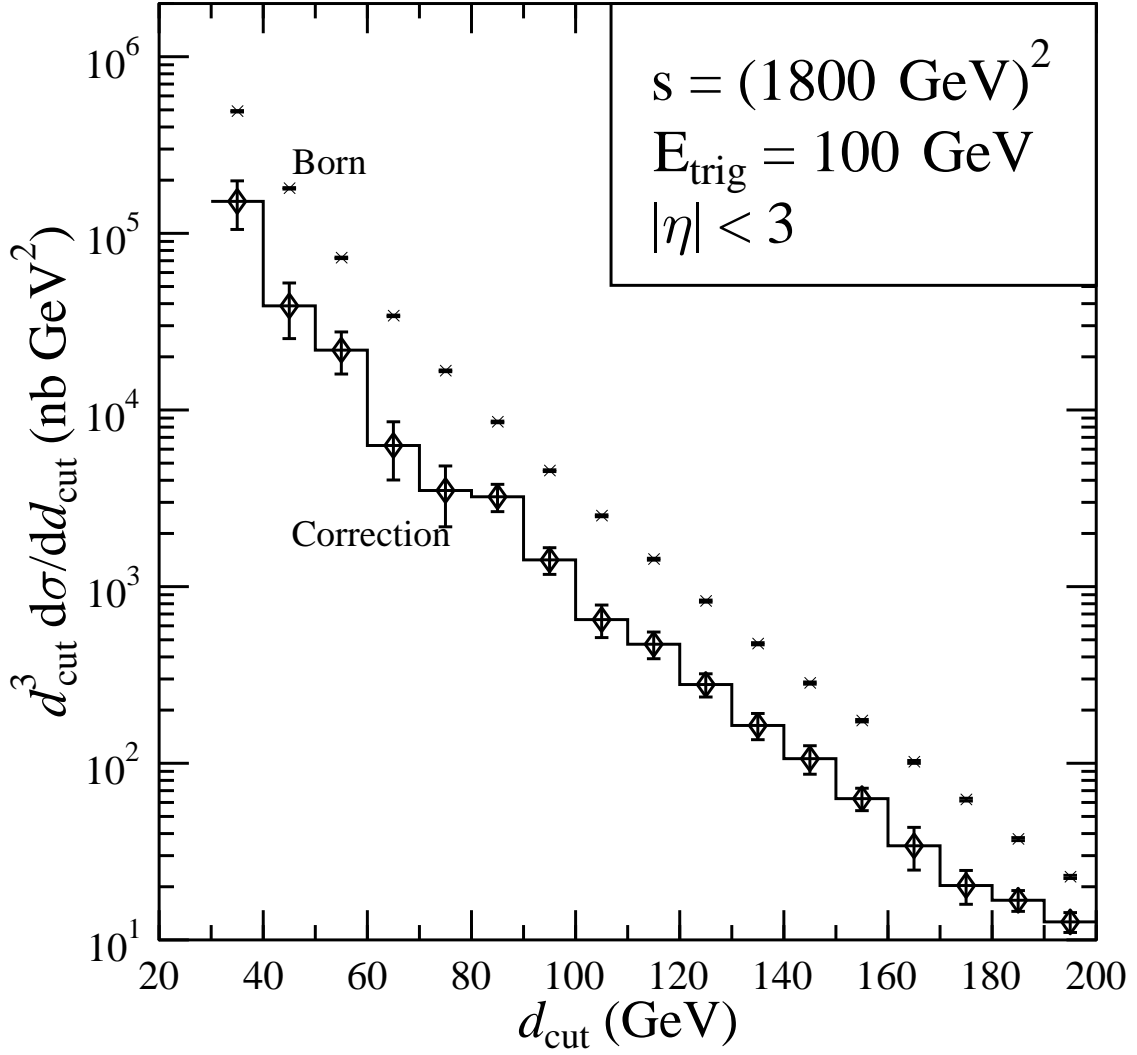}
\caption{Differential three-jet cross section $\dc^3\d\sigma/\d\dc$
vs $\dc$ for $\mu=\dc$ at Born level (crosses) and the higher order
correction to it (histogram). The errobars indicate the statistical
error of the Monte Carlo integration.}
\end{figure}

\end{document}